\documentstyle[aaspp,12pt]{article}
\begin{document}

\title{OBSERVATIONAL PROSPECTS FOR EXTRA-GALACTIC MICROLENSING EVENTS}
\author{Wesley N. Colley \footnote{Supported by the
Fannie and John Hertz Foundation, Livermore, CA 94551-5032}}
\affil{Princeton University Observatory, Princeton, NJ 08544--1001}

\begin{abstract}
We consider the feasibility of directly observing gravitational microlensing
in extra-galactic sources, whose stars are not generally resolved.  This
precludes use of the simple optical depth to microlensing formulation, which
is applicable only to resolved stars.  We, instead, extend this method
to consider observational constraints, such as seeing, sky background and
minimum detectable change in magnitude.  Our analysis provides
quantitative relations between these constraints and the expected observational
results, event duration and number of detections per year.
We find that an ambitious ground-based observer
should detect several events per year in M31.  We also consider detection
of microlensing in visual binary galaxies.  We find that, although these
objects may present hundreds of events per year, extremely short durations
would yield poor prospects for observation.

\end{abstract}

\section{INTRODUCTION}
	Gravitational lensing and its effects are well-known (Refsdal 1964).
However, meaningful astronomical results from this effect have been
daunting until recently.  A growing catalog of gravitationally lensed
high red-shift objects
assured us that the effect is observable, but until the last months of
1993, witnessing lensing by objects in our own galaxy had eluded us.

	In 1986, Paczy\'nski proposed that galactic events would be
detectable, but
the probability of seeing a star during an event was only $10^{-6}$, which
at that time seemed observationally prohibitive.  Nonetheless, as digital
detectors and faster computers became available,
realistic projects emerged
(Alcock, Axelrod, and Park 1989; Moniez 1990; Paczy\'nski 1991).
Now that these projects have detected several events
at a reasonable rate, we at last have direct observational
evidence of lensing in our galaxy.  (Alcock {\it et. al.}, 1993;
Aubourg {\it et. al.}, 1993; Udalski {\it et. al.}, 1994).

	But the most exciting aspect of this work is the implication for the
deflectors themselves.  Though dynamical evidence of a dark matter halo
in our galaxy was compelling (Binney \& Tremaine 1987),
no direct observations of this halo had been made.  The microlensing events
observed by the above-mentioned projects agree with a halo that
consists at least in part of 0.01$M_\odot$ - 0.1$M_\odot$ objects, direct
evidence at last of the dark matter halo, and important information on its
constitution.

	In this paper, we use such evidence for massive compact halo
objects (MACHO's) to
consider other possible observational targets for microlensing, particularly
very nearby galaxies and nearby visual binary galaxies.
Crotts (1992) proposed looking for lensing in M31, since we gain the
advantage of having both our halo and its halo to produce events.
Ballion, {\it et. al.} (1993) have
used Monte-Carlo methods to estimate event probabilities in this regime.
We address the problem analytically, with hopes to be quantitative
with respect to observational constraints.  Our analytical approach also
allows us to extend our range beyond M31.  Particularly,
if other galaxies have halos similar to ours, as one would expect from
dynamical evidence, observing a more distant galaxy through
the halo of a nearer one should produce microlensing events.

\section{THE MODEL}

	In estimating microlensing event probabilities, the seminal paper
by Paczy\'nski in 1986 covers the fundamental analysis thoroughly, and we
shall not repeat its bulk here, but as it is widely accepted, we will assume
its validity in our analysis.  However, the situation concerning us is
not quite the same one that faced him, so we offer a more specific analysis
of the extra-galactic regime.

	The most important difference between observing microlensing
events in galaxies, and observing them in the Galaxy or LMC, is that we
cannot necessarily resolve stars at the distances and stellar densities
involved, which is done readily in the Galaxy and at LMC.
Thus, to see a single event, the star must not only be
magnified significantly, but must be magnified sufficiently for
significant magnification of the seeing element containing it.  Figure 1
illustrates such a difference.  The lower dashed line is the stellar
flux, and the dotted curves show the time evolution of magnification of
this star (described thoroughly in Paczy\'nski 1991).  Now suppose
the star sits in a pixel whose magnitude
is two less than the star's (6.3$\times$ as bright, shown in figure 1
as unity).  For magnification of the pixel,
we must consider the {\it sum} of the lensed star's brightness and
the background's brightness, the evolution of which the solid curves show.
For instance, in the top curve, where the
star is magnified by nearly a factor of 10, the pixel
is magnified by only
a factor of two or so.  Figure 2 quantifies these results for typical
surface brightness and stellar magnitudes of very nearby galaxies, like M31.
Even a very bright star in M31 at the Einstein ring of a Galactic MACHO
would likely go undetected in a 17.5 magnitude background.  For more
typical stars, detection would be daunting even for high magnification events.
The observational requirements would thus
be far more demanding to see an event in such a regime.

	We must, therefore, reevaluate our probabilities of detection according
to these new observational constraints.  We first consider the familiar single
star
case, then examine how the probability will be modified by placing the star
in a bright background.  Vietri and Ostriker (1983) give the ubiquitous
equation for optical depth to microlensing which nominally describes the
instantaneous probability of seeing significant magnification of one star
against a black background.

\begin{equation}
P_* = {1\over {Area}}{\sum \pi R_{0,i}^2} =
 {\int{{4\pi G\rho D}\over {c^2}}}dD \equiv \tau
\end{equation}

	This equation states that the probability is simply the ratio of the
area inside the deflectors' Einstein radii, to the total area of the
deflector plane.
(If the reader is unfamiliar with the notation, he should consult Paczy\'nski
[1991], but here are
the basic definitions:

\begin{equation}
R_0^2 \equiv {{4Gm_d D}\over c^2}, ~
D \equiv {{D_d(D_s - D_d)}\over{D_s}}
\end{equation}

with $D_s$ = source distance, $D_d$ = deflector distance and $R_0$ = the
Einstein radius which is required proximity of a light ray to the deflector
for significant gravitational bending.)

	However, we have pointed out that an unresolved star requires more
substantial magnification than a resolved star, so we must modify the areas
in the sum accordingly.  Since more magnification means
smaller radii, and much less area, one may expect the probability for
the required magnifications to be prohibitively small.

	But consider Tonry's success at measuring small fluctuations
among pixels in galaxies (Tonry and Schneider 1988).  He measures tiny
fluctuations to estimate distances to galaxies quite
accurately.  Could we then employ a similar method for detecting lensing
events?  Or would Tonry fluctuations confuse our search?

	Figure 3 is a histogram plot of number of pixels vs. pixel
surface magnitude from a Monte-Carlo distribution of stars
in a 96 $\times$ 96 pixel
region of sky at $0.5''/pixel$ and $\sigma_{seeing} = 0.5''$ (FWHM = $1.18''$
with sky
background at $21^{mag}/(1'')^2$.  The three simulated star-fields,
with a luminosity
function detailed in section 3, represent regions of
surface magnitudes 20, 18.5, and 17.0 (factors of four in brightness from
right to left) in a galaxy at $1Mpc$.  Increasing the brightness
means increasing the number of stars per pixel, which should mean less
statistical fluctuation.  And moving from right to left, one sees the
expected sharpening of the distribution function.  The graph has two
implications
for our purposes.  The first is that even at low surface brightness, the
distribution is quite narrow, implying that events of moderate magnification
may be detectable above the pixel noise.
The second is that, like the statistical fluctuations, events
will be damped out as the pixel brightness increases, even as
the number of stars available for lensing increases.  We shall explore the
manifestation of these competing effects in detail in section 3.

	In any case, for a given apparatus, there is a minimum
observable change in magnitude, or detection threshold,
$\Delta m_0$.  Since we require several points to claim an event, we
define $\Delta m_0$ at the half-maximum magnification, $A_h$, which gives us
this simple expression.

\begin{equation}
{\Delta m_0 = -2.5log}{{A_h qf_* + f_{pix}}\over{qf_* + f_{pix}}}
\end{equation}

where the $f_*$ is the star's flux, $f_{pix}$ is the total flux of the pixel,
including the sky background,
and $q$ is the fraction of light in one pixel according to the point-spread
function.  With the required half magnification defined, we can find
the maximum radius which produces it.  Refsdal (1964) gives

\begin{equation}
A_h = {{u_h^2 + 2}\over{u_h\sqrt{u_h^2+4}}}
\end{equation}

where $u_h \equiv R_h/R_0$ is the dimensionless radius.

	So, at a given radius, (3) and (4) allows us to find the minimum
stellar flux [maximum magnitude, $m_V(u_h)$] required to
produce an event of $\Delta m_0$.  We now modify the
optical depth equation, so that instead of calculating the areas inside
the Einstein radius, we calculate an effective area in which the magnification
is sufficient to brighten the pixel by at least $\Delta m_0$.

	Consider differential annuli in radius, each with its own maximum
stellar magnitude, $m_V(u_h)$.  We weight the differential area of the annulus,
$2\pi u_hdu_h$, with with the relative number of stars brighter than
$m_V(u_h)$, according to the normalized luminosity function.

\begin{equation}
N(u_h) = \int_{m_V,min}^{m_V(u)} \phi(m_V^\prime) dm_V^\prime
\end{equation}

We now integrate over radius to find the effective area.

\begin{equation}
S_{eff} =
	\int_{u_{h,0}}^{u_{h,1}}
	2\pi u_h^\prime
	N(u_h^\prime) du_h^\prime
\end{equation}

	Since the brightest stars require the least magnification, and hence
allow the greatest radius, plugging the flux of the brightest stars into
equations (3) and (4) determines the maximum radius, $u_{h,1}$.
For the minimum radius, $u_{h,0}$, we
use the approximate minimum radius criterion for point-source treatment
given by Witt and Mao (1994) for a solar radius star in the source.  This
point-source criterion gives a radius significantly smaller than that of
the dimmest stars at the required magnification.  Since the criterion goes
inversely with magnification, as does the radius (for moderate to large
magnifications), we can be confident that the point-source treatment is valid.

	Plugging into the optical depth formula we obtain
the expected probability of a single star brightening a pixel by at least
$\Delta m_0$.

\begin{equation}
\tau_{eff} = {1\over{Area}}{\sum R_{0,i}^2 S_{eff}} =
	\int{{4 G\rho D}\over {c^2}}S_{eff} dD
\end{equation}

	By hypothesis, however, the number of stars per pixel is not usually
one, so we multiply this probability by the number of stars per pixel to obtain
the probability per pixel.

\begin{equation}
P_{pixel} = N_*\tau_{eff}
\end{equation}

with

\begin{equation}
N_* = {{f_0[\exp_{10}(-0.4\mu_V)-\exp_{10}(-0.4\mu_{V,sky})]}\over
	\langle f_* \rangle}
\end{equation}

which is simply the flux of stars in the pixel, divided by the
expectation value of the stellar flux.

	Armed with this probability, we still lack a practical estimate
for observational feasibility, because we have no idea how long an event
will last.  We thus commence analysis of event durations.

	So as not to offend dynamicists, I will not assume
any halo or disc model more complicated than estimating that deflectors
velocities are random at 200$km/s$ with respect to their parent galaxy.
With no attempt to straighten out order unity effects,
such as projections, or large effects such as inter-galactic motions,
we shall proceed.  (These results scale inversely with velocity, so
pundits may make corrections as they deem necessary.)

	As we know, stellar magnification depends only on radius (distance
from the light ray to the deflector).  A minimum
magnification, $A$, thus corresponds to a maximum radius, $u$,
within which all stars are magnified by at least $A$.  For
each circle of radius $u$, all stars whose impact parameter, $b$, is less than
$u$ will pass through the circle twice along their paths.
The path defines a chord on the circle, whose
length is simple to calculate.  Dividing by the velocity gives the crossing
time from one side of the circle to the other.

\begin{equation}
t_{cross}(b) = {{2\sqrt{r^2 - b^2}}\over v}
\end{equation}

	For our specific problem, $r = u_h$, and
$v = R_0^{-1} \cdot 200km/s$ gives the result appropriate to our
problem.  Recall that $u_h$ is the radius where the pixel is at
half-maximum magnification, so that $t_{cross}$ become the full-width at
half-maximum duration, $t_{fwhm}$.

	Since events have equal probability of occurring at all impact
parameters, we perform the weighted integral over impact parameter
to compute an expectation value for $t_{fwhm}$.
First, we find the maximum magnification as a
function of impact parameter using (4). ($A_{max}(b)$ and $b$ replace $A_h$ and
$u_h$).  The half-maximum magnification is $A_h(b) = [A_{max}(b)+1]/2$,
which we insert back into (4) to obtain the half-maximum radius, $u_h(b)$.
To find the weighting factor, $N(b)$, we use (3) with $A_h(b)$, which
gives the required stellar magnitude at each impact parameter.  As before,
the weighting factor is given by the integrated luminosity function.
The limits of integration, $b_0$ and $b_1$, correspond to the
half-maximum radius values as given for (6).

\begin{equation}
\langle t_{fwhm}\rangle(m)
 = {{2R_0(m)\int_{b_0}^{b_1}
	N(b)
	\sqrt{u_h^2 - b(u_h)^2}\;db}\over
	{\int_{b_0}^{b_1}
	N(b)db}}
	\equiv T \cdot R_0(m)
\end{equation}

	Furthermore, since $R_0$ is explicitly dependent on mass,
obtaining an expectation
value requires an integration of $t_{fwhm}(m)$ against the mass-function,
$\psi(m)$, which reduces to finding an expectation value of $\sqrt{m}$ (see
[2]).

\begin{equation}
\langle t_{fwhm}\rangle = T
	\int_{m_0}^{m_1} \sqrt{{4Gm^\prime D}\over{c^2}}
	\psi(m^\prime)dm^\prime
\end{equation}

	Having defined the probability of an event above the half-maximum
detection threshold, and the expected time of such an event, the expected
event frequency becomes $\langle P \rangle/\langle t_{fwhm} \rangle$.  For
useful units, we simply convert $\langle t_{fwhm} \rangle$ into years to find
the expected number of events per year.

\section{RESULTS}

	The equations in the previous section would be laborious to
integrate analytically, and even if it were possible, such an analysis would
limit flexibility in modifying the various parameters.  We thus employed
a fifth-order polynomial approximation to integrate them numerically
according the method found in Press {\it et. al.} (1988).

	Crucial to all results from our analysis is the large number of input
parameters, which in principle could be manipulated to give practically any
answer desired.  We thus designed our analysis to allow flexibility with
respect to these parameters, and if one thought our nominal values
na{\"\i}ve, they could be easily changed to better values.

\bigskip
{\bf 3a. M31}
\bigskip

	First, we consider prospects for M31.  As pointed out by Crotts (1992),
halo objects of either M31, or the Galaxy could provide lensing.  For now, we
study only the effect of Galactic deflectors; we will consider the
effect of M31's halo on optical depth later.  However, we shall presently
explain why the event durations will be nearly identical for both cases.
Witt and Mao (1994) give the explicit expression for duration,
including source motion, deflector motion and observer motion.  If one
assumes that the velocities and distances of the Galactic deflectors relative
to Earth are comparable to those of M31's deflectors relative to its
stars, he can readily show that the two cases give the
same duration to within a factor of
$(1 \pm \delta v_{s,o}/v_d)$, with $|\delta| \lesssim 10kpc/770kpc$ and
$v_{s,o}$, the relative velocity between the source and observer, of
order $v_d$.
Perhaps counter-intuitive at first, because of the slower angular motion
of M31's deflectors, the angular Einstein radius at M31 is also less by
that same factor.

	To estimate the parameters of the M31 case,
we quite simply set the optical depth to $10^{-6}$,
and the deflector mass
to $0.02M_{\odot}$.  To emulate the stars of M31, we use the Hodge (1988)
luminosity function, $\phi(M_V) \propto 10^{0.57M_V}$ over a range of
$-4 < M_V < 8$, which covers early B stars down to late K stars (Mihalas \&
Binney 1981).  To estimate event duration, we set the deflectors
in motion about both galaxies at $v = 200km/s$.

	Setting $\Delta m_0 = -0.1$ and FWHM seeing = $1''$ gives figures
4(a) and 4(b), which show our results for event probability
per CCD ($2048^2 \times P_{pixel}$; pixel width = $0.44''$), and expected
event duration in hours.  The ordinate of figure 4(a)
is the integrand in equation (6), just as the ordinate of 4(b) is the integrand
in equation (10).  Plotting this way allows one to see
where the bulk of the contribution for probability and duration lies, and
by definition, gives the final expected values as the areas under the curves.
Both figures plot two
families of curves labelled in surface magnitude.  The solid family is
for $\phi(M_V) \propto 10^{0.57M_V}$, the dotted family for
$\phi(M_V) \propto 10^{0.40M_V}$.

	Evaluating 4(a), we consider that at small radius
(dimensionless radius at half-maximum magnification, $u_h$),
the numerous dim stars will produce sufficient
pixel magnification, but at large radius, only the few
bright stars will produce sufficient magnification.  However, the contribution
to area is much greater for the high impact parameters, so it is hard
to guess which effect will dominate.  Figure 4(a) immediately resolves
the confusion; the mild lensing of bright stars contributes the bulk of the
probability, in agreement with Baillon {\it et. al.} (1993).

	The shape of the dotted curves can be understood quite easily.
Using 0.4 as the coefficient of $m_V$ in the magnitude function puts the
luminosity function inversely proportional to flux, while for large
magnifications the required flux goes linearly with impact
parameter (see Section 2), and the two cancel.  The extra factor of
impact parameter from the logarithmic integration generates the linear
slope of the curves, which turn over when the luminosity function runs
out of stars on the bright end.

	Examining the family of curves, we see that for low surface magnitudes
(bright backgrounds) the event duration drops dramatically.  Since the
probability scales with the area under these curves (in linear space),
the probability drops in the same way.
Although putting more stars in the field by
turning up the brightness increases the number
of stars available for lensing, it more sharply stiffens the magnification
requirements.  In fact, observing a region at
$\mu_V = 17.5^{mag}/(1'')$ increases event probability by more than a
factor of ten compared to observing in a 14.5 magnitude region.
Notice, however, a second effect comes in at the
very dim end.  Here the $21^{st}$ magnitude sky has become comparable
to the brightness in stars, so that fewer and fewer stars contribute
to the total brightness.  With less and less stars, but the same pixel
brightness
to overcome, the probability drops off quickly at the dim end.

	In figure 4(b), the shapes of the curves are much the same as for 4(a),
because as with probability, there is the luminosity function factor and
two factors linear in $b$, the dimensionless impact parameter.
Notice that again the
curves roll over sooner for the brighter backgrounds, indicating longer
expected durations for dimmer regions, until the sky background is reached.
This faint regime contains the only significant difference between the
figures.  In 4(b), the sky changes the separation of the curves along
the linear direction, so
that the expected duration approaches an asymptotic value at the faint end,
instead of
dropping away altogether as does the probability.  This is to be expected,
because the number of stars lensed has nothing to do with the
durations of the events.

	Fixing $\mu_V$ at a constant value, we examine the effects
of seeing and limiting detection threshold, $\Delta m_0$.
In figures 5(a), (b) and (c), we have plotted respectively the
event probability per CCD, the expected event duration, and
event number per year, as functions of seeing, for several values of
$\Delta m_0$.  On each graph two values of $\mu_V$ are used, 19.0 (solid lines)
and 17.5 (dotted lines).  For surface magnitudes greater than about 19,
the $21^{st}$ magnitude sky begins to interfere and decreases the
probability substantially (recall figure 4[a]).

	Figure 5(a) shows that event probability decreases
by more than two orders of magnitude from perfect
seeing to a seeing disc of a $2''$ diameter.  Similarly, the curve families
indicate that large detection thresholds hinder event detection;
between $\Delta m_0 = -0.1$ and $\Delta m_0 = -0.5$, the probability drops
by a factor of order 25.

	As we move to the event durations, as seen in figure 5(b), we may
wonder why duration depends on seeing at all.  The durations of the stellar
events are surely independent of observation, but the probability of seeing the
longer, lower magnification events drops with seeing, so the expected event
duration decreases with seeing, but not sharply.  From
perfect seeing to a seeing disc of $2''$, the expected duration decreases
by about a factor of order 10, and as with the probability, event
duration decreases with poorer detection
thresholds, by a factor of order five.  Ostensibly, good skies and sensitive
detectors are of the essence for event detection.

	Dividing probability by duration gives us the important quantity of
number of events per year in figure 5(c).  Together with 5(b), these plots
probably provide the most useful reference for observers, since they give
expected observational results as functions of observational criteria.
{}From these graphs, we see that for reasonable seeing, and detection
threshold of 0.1 magnitudes, about ten three-hour events will occur
each year (for a $2048^2$ CCD, $15' \times 15'$ field),
which is probably too few to observe.
However, Crotts (1992) points out that observing the far side of M31, through
the bulk of its halo, increases optical depth by a factor
of ten, which increases the number of events by the same factor.  With good
seeing, and ability to detect 0.1 magnitude changes, an observer
may expect of order 100 events per year.  Thus, if one observes on 120 nights,
with six hours of data per night, he should see several events.
Although somewhat more pessimistic than Baillon {\it et. al.} (1993),
who estimate 50 detections per year, our results indicate a tenable
number as well.

	Having considered the effects of sky background, seeing, surface
magnitude and detection threshold explicitly, we propose that microlensing
would be detected in M31 under reasonable observational constraints.
A particular method is that of Tomaney \& Crotts (1994), who
have demonstrated efficacy in detecting unresolved variable stars in
M31.  We submit that using a similar method would produce microlensing event
detections (as Crotts [1992] points out).  Our results should optimize such a
search in both target region and exposure schedule.  Namely, we find
that the most productive regions of M31 will be the far side of disc in areas
about a magnitude brighter than the sky background.
However, large light gathering power, steady skies, and detectors capable of
several exposures per hour, will be critical in detecting the bulk
of events.

\bigskip
{\bf 3b. Nearby Visual Binary Galaxies}
\bigskip

	With our analytical method, we have the added benefit of extending
our range of objects beyond M31.  Particularly, we can determine observational
prospects for lensing events in visual binary galaxies.  Figure 6(c)
demonstrates that the number of events per year actually increases as the
distance to the source increases, and thus shows the need to
explore carefully the idea of observing microlensing beyond M31.

	The main advantage with these sources is that instead of using
the halos of our galaxy or the source galaxy, we use the halo of an
intermediate
galaxy to supply the deflectors.  As equation 2 indicates, the geometry of
lensing favors deflectors near the midpoint between the observer and the
source.  In figures 6(a), (b) and (c), instead of holding the optical depth
constant, we have
held the surface mass density in the deflector constant at $\Sigma = 200
M_\odot/pc^2$, and placed the deflector at half the source distance.  We have
labelled the five curves in surface magnitude, while setting the other
parameters constant:  $\Delta m_0 = -0.1$, FWHM seeing $= 1''$, and
all other parameters are as in section 3a.

	Although figure 6(c) presents an initially encouraging result, a
quick glance a figure 6(b)
shows that the durations could be prohibitively small at large distances.
To make matters worse, $200km/s$ could be a drastic underestimate of the
relative velocities in the deflector plane.  Since the source, deflector
and observer could all have motions at $\gtrsim 500km/s$, the durations
could, in fact, be much less.
Nonetheless, as I cannot estimate the light-gathering power of future
telescopes, nor the speed of future detectors, I shall proceed to
interpret our results briefly, while
keeping with the very optimistic $200km/s$ relative velocity in the
deflector plane.

	Figure 6(a) shows the probability per CCD that an event is underway
at a given time.  At very small distances, we can resolve stars, and we
recover the results for the resolved star case, where the probability increases
as the third power of distance.  Two of these factors are due to our
constant solid angle pixels.  The number surface density in stars must
increase as distance squared for maintain constant surface
magnitude, so over constant solid angle, the number of stars increases
with distance squared.  The
third factor is the linear increase in optical depth from equation (2).
Each curve, however, flattens when resolving stars becomes difficult due
to the seeing disc.  For the very bright regions, this happens quickly, but
the peak for a $\mu_V = 20.5$ extends to around $1Mpc$.  Beyond the peak,
the probability falls off linearly, which can be understood as follows.
The ratio of stellar flux to the constant pixel flux falls off as
distance squared, so the required magnifications must increase with this
factor for large magnification.
Equation (4) shows that in this limit, radius goes as
$A^{-1}$, so that $S_{eff}$ must go as $A^{-2}$.
Having lost four powers of distance from our original three,
we now understand the inverse
relation of probability and
distance.  One further note on probability is that galaxies become smaller
than the coverage of a single CCD chip, which will decrease the probability
by distance squared once that limit is reached.

	In figure 6(b), we see that the event duration converges to
$\sqrt{D_s}$ for all $\mu_V$ as distance goes to zero.  This result is expected
from equations 2 and 11.  The first departure from the convergent value is
where the seeing discs first begin overlap, so that the required magnifications
start rising.  From this point to the turn-over, the curve depends on the
luminosity function as bright stars are still effectively resolved, but the
dim stars are slipping into oblivion.  At the turn-over, even bright stars
are no
longer resolved, and the required magnifications grow rapidly.  As with the
probability, the required magnification increases with distance squared,
and the required impact parameters (thus event durations) go inversely with
magnification.  Thus
at large distances, duration contains two reciprocal factors of distance,
but since the square root factor remains, the duration falls off
with $D_s^{3/2}$.

	Figure 6(c) is easily understood then, as simply the quotient of
figure 6(a) and 6(b), so that in the end, the number of events scales with
the square root of distance.

	Considering figure 6(b) quantitatively, we see that
out to about a megaparsec, the duration slowly decreases from a few days to
several hours, but beyond a megaparsec, duration
decreases rapidly, down to minutes at $100Mpc$.  If fast enough detectors
and abundant photons existed,
there would be hundreds of events detected per year, but with
today's technology, we
require that events be about an hour in length for multiple exposures
per event.

	Since duration scales with the square root of deflector
mass, we may look to increase durations by increasing the deflector masses.
We have used $0.02M_\odot$ deflectors, as before, so in order to increase the
duration by a factor of ten, we would need to increase our minimum mass
to $1M_\odot$, an unlikely prospect for halo
objects.

	We then ask if we could only use the brightest stars in the luminosity
function, which have the longest possible duration.  Returning to figure 4(b),
however, we see that since the bulk of the expected
duration already lies in the high duration end (lensing of the brightest
stars), leaving little room for improvement in the luminosity function.

	The only hope then, is to employ the very high mass end of the
mass function.
To get to $1M_\odot$, one suffers a factor $10^{-4}$ in probability, which
is coupled with the factor of $10^{-1}$ due to the increased duration, to
give a factor of $10^{-5}$ in event rate.

	 Space-based observation may offer some help, however.
Figures 5(b) and 5(c) illustrate how the number of events increases
by a factor of more than five for space-telescope seeing vs. ground-based
seeing, while the duration increases by a factor of about two.  These
numbers still imply less than an event per year of one-hour duration.
When we recall that our drastic underestimate
of velocities pushes the durations up substantially, the picture is
bleak indeed.

	Nonetheless, for quantitative results, we did examine several
actual cases.  The tedious work of finding visual binaries within $100Mpc$, was
greatly relieved by the kindness of Marc Postman who allowed us
to peruse his unpublished data on such objects.  Since distances are poorly
known at this range, we assumed Hubble distances with $H_0 = 75km/s/Mpc$.
Our search uncovered many candidate pairs, the best of which
have $D_s \sim 50Mpc$, and $D_d \sim 10Mpc$.  Such pairs
give results similar to those mentioned above for both ground
and space-based projects.  Unless we were able to take many plates per
minute, the prospects seem poor for observing microlensing events in this
realm.

\section{Conclusions}

	We have presented an analytical study to determine the observational
prospects of observing microlensing in extra-galactic sources.  We estimate
that with reasonable ground-based equipment, several microlensing events will
be found in M31 over the span of a year.
Furthermore, we give detailed information on probabilities and durations of
observed events according to a range of observational constraints.  These
data should help to focus an observing project onto the correct region of
M31 according to equipment.  In general, we find that the far side of M31's
disc, in regions about a magnitude above the sky background, hold the most
promise for event detection.

	Nearby visual binary galaxies may also display individual
microlensing events.  However, even with space-based facilities, one
would require many plates per minute, an unlikely prospect at the present.

\acknowledgements{The author is most grateful to Bohdan Paczy\'nski, who not
only proposed the idea for pursuing this project, but also
provided critical guidance and advice through the duration of the project.
Thanks also
to Krzysztof Stanek and Hans Witt for their useful conversations, and to
Marc Postman for access to his unpublished catalog of visual binary galaxies.
We are grateful for NSF grants AST-9216494 and AST-9313620, which
provided funding for this project.}

\newpage

\begin{figure}
\begin{center}
{\bf FIGURE CAPTIONS}
\end{center}

\caption{Magnification of a pixel as a function of time in a
microlensing event.  The heavy
solid line is the total flux of the background.  The dashed line is the
relative flux of the star two magnitudes below the background.
The dotted curves are the stellar magnification
for impact parameters of $0.10R_0$, $0.25R_0$ and $0.40R_0$.  The solid curves
are the pixel magnification for the same impact parameters.  The squares
represent the half-maximum magnifications for each curve.  Notice that even
when the star is magnified nearly ten times, the pixel is only magnified by
a factor of 2.}

\caption{Magnification of pixel as a function of radius in a
microlensing event.  Each curve is labelled according to the stellar magnitude
of the lensed star.  The pixel background is at magnitude 17.5.
The brightest star has $M_V = -3.2$ at M31.  Events with bright stars at the
Einstein radius, and dim stars $0.01R_0$, would likely go undetected.}

\caption{Surface brightness fluctuations at $0.7Mpc$.  With a stellar
luminosity function of $\phi(M_V) \propto 10^{0.57M_V}$, $-4 < M_V < 4$ and
$\mu_{V,sky} = 21$, we have
plotted the distributions in pixel magnitudes for $\mu_V = $ 20, 18.5 and 17.
Since
lower surface magnitudes require more stars per pixel, the width of the
distribution decreases.}

\caption{(a) Differential event probability per CCD with respect to logarithmic
radius at half-maximum magnification, and (b) differential expected event
duration with respect to impact parameter,
for Galactic MACHO lensing of M31 stars (see text
for details).  The curves are labelled in surface magnitudes.  The solid
curves are for $\phi(M_V) \propto 10^{0.57M_V}$, the dotted curves for
$\phi(M_V) \propto 10^{0.4M_V}$.  The duration increases for lower surface
brightness, as does total probability.  The sky is at magnitude 21.  These
curves indicate that the bulk of duration and probability occurs in low
magnification lensing of high luminosity stars.}
\end{figure}

\newpage
\begin{figure}
\caption{(a) Event probability per CCD, (b) FWHM event duration, and (c)
number of events per year, as functions of seeing, for Galactic MACHO
lensing of M31 stars.  The
curves are labelled in detection threshold $(\Delta m_0)$.  The solid curves
are for $\mu_V = 17.5$, the dotted curves for $\mu_V = 20.5$.}

\caption{(a) Event probability per CCD, (b) FWHM event duration, and (c)
number of events per year,
as functions of source distance.  The curves are labelled in surface magnitude
in the source.  The deflectors are in a galaxy half-way between the observer
and the source ($D_d = D_s \div 2$).}
\end{figure}
\end{document}